
\def\data{\number\day.\number\month.\number\year}
\def\datasotto{\@datatrue\footline={\hfil{\rm \data}\hfil}}
\def\frac#1#2{{#1\over {#2}}}
\def\te#1#2{\Theta^{#1 #2}}

\def\diff#1#2{{\partial
#1\over\partial #2}}

\def\2est#1#2{dx^{#1}\wedge dx^{#2}}
\def\3est{dx^1\wedge dx^2\wedge dx^3}

\def\4est{dx^0\wedge dx^1\wedge dx^2\wedge dx^3 }

\def\0est#1#2{d#1^{#2_0}\wedge ...\wedge d#1^{#2_k}}

\def\esterno3{d\xi ^1\wedge d\xi ^2\wedge d\xi ^3}
\def\pretitolo{\par\vskip .7cm\noindent}
\def\postitolo{\par\nobreak\vskip .3cm\nobreak}
\def\capo{\par\noindent }

\def\menog{\sqrt {-g}}

\def\due{{1\over 2}}

\def\quadratino{\vbox{\hrule\hbox{\vrule\vbox to 7 pt
{\vfill\hbox to 7 pt {\hfill\hfill}\vfill}\vrule}\hrule}}

\def\charslash#1{\setbox2=\hbox{$#1$}
\dimen2=\wd2
\setbox1=\hbox{/}\dimen1=\wd1
\ifdim\dimen2>\dimen1
\rlap{\hbox to \dimen2{\hfil /\hfil}} #1
\else
\rlap{\hbox to \dimen1{\hfil$#1$\hfil}}
/ \fil\fi}
\def\tagliato#1{\setbox2=\hbox{$#1$}
\dimen2=\wd2
\setbox1=\hbox{{--}}\dimen1=\wd1
\ifdim\dimen2>\dimen1
\rlap{\hbox to \dimen2{\hfil{--}\hfil}}
#1
\else
\rlap{\hbox to \dimen1{\hfil$#1$\hfil}}
{{--}}
\fil\fi}
\font\strana=cmti10
\def\L{\hbox{\strana \char'44}}

\def\dib#1{\partial _{#1}}

\def\nadib#1{\nabla _{#1}}

\def\gi#1#2{g_{#1#2}}

\def\gia#1#2{g^{#1 #2}}

\def\ixi#1#2{\xi^{#1 }_{\ #2 }}

\def\h#1#2{h_{#1#2}}
\def\id#1#2{\delta ^{#1}_{#2}}

\catcode`@=11
\def\b@lank{ }

\newif\if@simboli
\newif\if@riferimenti
\newif\if@incima
\newif\if@bozze

\def\bozze{\@bozzetrue
\immediate\write16{!!! INSERISCE NOME EQUAZIONI !!!}}
\newwrite\file@simboli
\def\simboli{
\immediate\write16{ !!! Genera il file \jobname.SMB }
\@simbolitrue\immediate\openout\file@simboli=\jobname.smb
\immediate\write\file@simboli{Simboli di \jobname}}

\newwrite\file@ausiliario
\def\riferimentifuturi{
\immediate\write16{ !!! Genera il file \jobname.AUX }
\@riferimentitrue\openin1 \jobname.aux
\ifeof1\relax\else\closein1\relax\input\jobname.aux\fi
\immediate\openout\file@ausiliario=\jobname.aux}

\newcount\eq@num\global\eq@num=0
\newcount\sect@num\global\sect@num=0
\newcount\lemm@num\global\lemm@num=0

\newif\if@ndoppia
\def\numerazionedoppia{\@ndoppiatrue\gdef\la@sezionecorrente{
\the\sect@num}}

\def\se@indefinito#1{\expandafter\ifx\csname#1\endcsname\relax}
\def\spo@glia#1>{}

\newif\if@primasezione
\@primasezionetrue

\def\s@ection#1\par{\immediate
\write16{#1}\if@primasezione\global\@primasezionefalse\else\goodbreak

\vskip\spaziosoprasez\fi\noindent
{\bf#1}\nobreak\vskip\spaziosottosez\nobreak\noindent} %

\def\sezpreset#1{\global\sect@num=#1
\immediate\write16{ !!! sez-preset = #1 } }

\def\spaziosoprasez{50pt plus 60pt}
\def\spaziosottosez{15pt}

\def\sref#1{\se@indefinito{@s@#1}\immediate\write16{ ???
\string\sref{#1}
non definita !!!}
\expandafter\xdef\csname @s@#1\endcsname{??}\fi\csname
@s@#1\endcsname}

\def\autosez#1#2\par{
\global\advance\sect@num by 1\if@ndoppia\global\eq@num=0\fi
\global\lemm@num=0
\xdef\la@sezionecorrente{\the\sect@num}
\def\usa@getta{1}\se@indefinito{@s@#1}\def\usa@getta{2}\fi
\expandafter\ifx\csname @s@#1\endcsname\la@sezionecorrente\def
\usa@getta{2}\fi
\ifodd\usa@getta\immediate\write16
{ ??? possibili riferimenti errati a \string\sref{#1} !!!}\fi
\expandafter\xdef\csname @s@#1\endcsname{\la@sezionecorrente}
\immediate\write16{\la@sezionecorrente. #2} \if@simboli
\immediate\write\file@simboli{ }\immediate\write\file@simboli{ }
\immediate\write\file@simboli{ Sezione
\la@sezionecorrente : sref. #1}
\immediate\write\file@simboli{ } \fi
\if@riferimenti
\immediate\write\file@ausiliario{\string\expandafter\string\edef
\string\csname\b@lank
@s@#1\string\endcsname{\la@sezionecorrente}}\fi
\goodbreak\vskip 48pt plus 60pt
\noindent{\bf\the\sect@num.\quad #2}
\if@bozze
{\tt #1}\fi
\par\nobreak\vskip 15pt
\nobreak\noindent}

\def\semiautosez#1#2\par{
\gdef\la@sezionecorrente{#1}\if@ndoppia\global\eq@num=0\fi
\if@simboli
\immediate\write\file@simboli{ }\immediate\write\file@simboli{ }
\immediate\write\file@simboli{ Sezione ** : sref.
\expandafter\spo@glia\meaning\la@sezionecorrente}
\immediate\write\file@simboli{ }\fi
\s@ection#2\par}


\def\eqpreset#1{\global\eq@num=#1
\immediate\write16{ !!! eq-preset = #1 }	}

\def\eqlabel#1{\global\advance\eq@num by 1
\if@ndoppia\xdef\il@numero{(\la@sezionecorrente.\the\eq@num)}
\else\xdef\il@numero{(\the\eq@num)}\fi
\def\usa@getta{1}\se@indefinito{@eq@#1}\def\usa@getta{2}\fi
\expandafter\ifx\csname
@eq@#1\endcsname\il@numero\def\usa@getta{2}\fi
\ifodd\usa@getta\immediate\write16 { ??? possibili riferimenti
errati a
\string\eqref{#1} !!!}\fi \expandafter\xdef\csname
@eq@#1\endcsname{\il@numero} \if@ndoppia
\def\usa@getta{\expandafter\spo@glia\meaning
\la@sezionecorrente.\the\eq@num}
\else\def\usa@getta{\the\eq@num}\fi
\if@simboli
\immediate\write\file@simboli{ Equazione
\usa@getta : eqref. #1}\fi
\if@riferimenti
\immediate\write\file@ausiliario{\string\expandafter\string\edef
\string\csname\b@lank @eq@#1\string\endcsname{\usa@getta}}\fi}

\def\eqref#1{\se@indefinito{@eq@#1}
\immediate\write16{ ??? \string\eqref{#1} non definita !!!}
\if@riferimenti\relax
\else\eqlabel{#1} ???\fi
\fi\csname @eq@#1\endcsname }

\def\autoeqno#1{\eqlabel{#1}\eqno\csname @eq@#1\endcsname\if@bozze
{\tt #1}\else\relax\fi}
\def\autoleqno#1{\eqlabel{#1}\leqno(\csname @eq@#1\endcsname)}

\def\titoli#1{\@incimatrue\nopagenumbers\xdef\prima@riga{#1}
\if@incima
\voffset=+30pt
\headline={\if\pageno=1{\hfil}\else\hfil{\sl
\prima@riga}\hfil\folio\fi}
\fi}

\catcode`@=12

\hsize 12cm
\magnification 1200

\hsize 17truecm
\vsize 24truecm
\font\ten=cmbx10 at 13pt
\font\twelve=cmbx10
\font\eight=cmr8

{
\centerline{{\twelve Centre de Physique
Th\'eorique}\footnote{$^{\star}$}{\eight Unit\'e Propre de Recherche
7061}{\twelve - CNRS - Luminy, Case 907}}
\centerline{{\twelve F--13288 Marseille Cedex 9 -
France}}

\vskip 4truecm

\centerline{\ten RELATIVISTIC ELASTOMECHANICS IS A }
\centerline{\ten GAUGE--TYPE THEORY}
\bigskip

\centerline{  {\bf Jerzy KIJOWSKI}\footnote{ $^1$}{\eight Permanent
address : Centre for Theoretical Physics, Polish Academy of
Sciences}\footnote{}{\eight Al. Lotnik\'ow 32/46; 02--668 Warsaw,
Poland } {\bf and Giulio MAGLI}\footnote{ $^2$}
{\eight Permanent address : Dipartimento di Matematica del
Politecnico di  Milano}\footnote{}{\eight  Piazza Leonardo da Vinci
32, 20133 Milano, Italy }  }

\vskip 2truecm

\centerline{\bf Abstract}

\medskip

A new approach to relativistic elasticity theory is proposed.
In this approach the theory becomes a gauge--type theory, with the
diffeomorphisms of the material space playing the role of gauge
transformations. The dynamics of the elastic material is expressed
in
terms of three independent, hyperbolic, second order partial
differential
equations imposed on three (independent) gauge potentials. The
relationship with the Carter-Quintana approach is discussed.

\vskip 3truecm

\noindent Key-Words : relativistic elasticity, variational
principles, gauge theories.
\bigskip

\noindent November 1994

\noindent CPT-94/P.3102

\bigskip

\noindent anonymous ftp or gopher : cpt.univ-mrs.fr

\footline={}

\vfill\eject    }

\titoli{J. Kijowski, G. Magli, Relativistic elastomechanics is a
gauge--type theory}
\capo
{\bf 1. Introduction}
\postitolo

The interaction between
the gravitational field and an elastic solid body became quite an
important problem in astrophysical applications, since the
discovery that
the crust of neutron stars probably exists in the form of a solid,
due to
the process of crystallization of dense neutron matter (see e.g.
McDermott, Van Horn \& Hansen 1988). Recently, the present authors
proposed a new approach to relativistic elasticity (Kijowski \&
Magli
1992). The aim of the present paper is to show the relationship
between
our approach and the theory proposed by Carter and Quintana (1972)
and
developed in Carter (1980). It turns out that the method of
deriving the
dynamical equations of the theory used by the latter authors is
analogous
to the derivation of (special-relativistic) Maxwell equations {\it
via}
the (general-relativistic) Hilbert variational principle, without
introducing the notion of electromagnetic potential. On the other
hand,
our approach is analogous to the standard description of
electrodynamics
in terms of potentials.\par In our formulation all the physical
quantities (like e.g. the stress and the strain tensors, the matter
current and so on) are defined in terms of first order derivatives
of the
potentials. This way, all the compatibility conditions of the
theory are
automatically satisfied. As a consequence, the dynamics can be
formulated
in terms of three (independent) second-order hyperbolic partial
differential equations imposed on three (independent) unknown
functions:
the gauge potentials. This simplifies considerably the dynamical
structure of the theory.
\par
The equivalence between the two formulations is a straightforward
consequence of the following,
remarkable feature of the relativistic mechanics of continuous
media:
the symmetric and the canonical
energy--momentum tensors of the theory do coincide (due to the
convention which is generally used, they actually coincide up to a
sign -
see e.g. Jezierski \& Kijowski 1991).
It is well known that this is not the case for a general
relativistic
field theory, like e.g. electromagnetism, although the relationship
between the two tensors is well
understood (see e.g. Kijowski \& Tulczyjew 1979). In the particular

case of relativistic elasticity, the simplest way to understand the
equivalence between the two tensors is based on the following
observation. Relativistic elasticity interacting with the
gravitational
field may be regarded as the theory of two symmetric tensor fields:
the
physical metric $g$ and the {\it material metric} $h$ (see section
2). As
will be seen in the sequel, the Lagrangian of the theory depends on
both
metrics {\it via} their combination $g^{\mu\nu}h_{\nu\lambda}$
only.
Hence, its variations with respect to both metrics coincide (up to
a
sign).

\pretitolo
{\bf 2. Kinematics}
\postitolo

Let ${\cal M}$ be the general--relativistic space--time equipped
with
a pseudo--rie\-man\-nian metric tensor $g_{\mu \nu }$ ($\mu ,\nu
=0,1,2,3$) of signature $(-,+,+,+)$. Suppose that ${\cal M}$ (or an
open
domain ${\cal O}\subset {\cal M}$) is filled with a continuous
material.
The collection of all the trajectories of the material points may
be
considered as a 1-dimensional foliation of ${\cal O}$. This
foliation can
be described, as usual, by a normalized, time-like, future oriented
vector field $u^\mu$, tangent to the trajectories and called the
{\it
four-velocity}: $$ u^\mu u_\mu =-1 \ .\autoeqno{norma}
$$
To complete the description of the configuration of the material
one has
 to include the information about such quantities as the matter
density,
internal strains, temperature etc., according
to the physical character of the material. In the case of a
barotropic
and isotropic elastic body, which we are going to describe in the
present
paper, the complete information about the configuration is carried
by its
internal {\it material metric}. It describes the ``would be''
rest-frame
distances between adjacent molecules of the body, if the
corresponding
infinitesimal portion of the body had been extracted from the rest
of the
material and left in the perfectly relaxed state. Mathematically,
the
material metric will be described by a symmetric tensor field
$h_{\mu\nu}$.

The ``material distance'' between points belonging to the same
trajectory has to vanish, because they correspond to the same
molecule of
the body. Hence, $h_{\mu\nu}$ has to be orthogonal to the velocity:
$$
h_{\mu \nu} u^\mu =0 \ . \autoeqno{orto} $$ This implies that $h$
has the
signature $(0,+,+,+)$, and therefore it may assume the value of any
non-negative, symmetric tensor, having precisely {\it one}
vanishing
eigenvalue, and such that the corresponding eigenvector is
time-like. At
each point of ${\cal M}$ there is a 9-parameter family of such
objects.
Observe that the information about the velocity $u$ is already
contained
in $h$. Indeed, $u$ may be defined as the unique time-like,
future-oriented, normalized eigenvector of $h$. We conclude that
$h$
carries the entire information about the configuration of the
physical
system in question. The dynamical equations of relativistic
elasticity
theory may be formulated as first order differential equations
imposed on
the quantity $h$.

We are going to describe only materials without memory, i.e. such
that
the material distance
between two adjacent particles
remains constant during the evolution. More precisely, this
condition
 means the following:
extracting an infinitesimal portion
of the material and letting it relax leads to the same distance
between
 the particles, independently of the moment at which
the portion has been extracted.
Mathematically, this
means that the metric
$h$ is ``frozen'' in the material, i.e. its Lie derivative with
respect
 to $u$ vanishes:
$$
\L_{u} \h \mu\nu = 0 \ , \autoeqno{lie}
$$
where
$$
\eqalign{
\L_{u} \h \mu\nu := & u^\lambda \nadib \lambda \h \mu\nu + \h
\mu\lambda
 \nadib \nu u^\lambda + \h \nu\lambda \nadib \mu u^\lambda = \cr
= &
u^\lambda \dib \lambda \h \mu\nu +
\h \mu\lambda \dib \nu u^\lambda + \h \nu\lambda \dib \mu u^\lambda
 \ . }\autoeqno{lh}
$$
As $u$ is also a function of $h$, equation \eqref{lie} has to be
considered as an identity imposed on $h$ alone; only those $h$
describe
physically admissible configurations of the material, which fulfill
this
condition. Due to the identity $$ u^\mu \L_u \h \mu\nu =\L_u (u^\mu
\h
\mu\nu )-\h \mu\nu \L_u u^\mu \equiv 0 \ ,
$$
there are only 6 independent conditions in \eqref{lie}, imposed on
the 9 independent components of $h$. We conclude that the
configurations
of an elastic material are described by 3 independent functions
(degrees
of freedom) defined implicitly by the identities \eqref{lie}.

\pretitolo
{\bf 3. Relativistic strain tensor}
\postitolo

The material is locally relaxed at a point $x\in {\cal M}$ if and
only if the physical distances described by the metric $g$ coincide
with
the material distances described by $h$. This happens if the
material
metric coincides with the physical metric on the subspace
orthogonal to
$u$, i.e.~if the following equation is satisfied at $x$: $$ h_{\mu
\nu} =
E_{\mu\nu} \ , \autoeqno{relax} $$ where by $E$ we denote the
orthogonal
projector $$ E_{\mu\nu} :=\gi \mu\nu +u_\mu u_\nu \ . $$ In a
generic
situation
$h$ is not equal to $E$.
The bigger is the difference between them, the stronger is the
state of
strain of the material. There are many ways to measure this state;
one
possibility is to introduce their difference (Cattaneo 1973, Maugin
1978): $$
\Sigma_{\mu\nu}:=\due (E_{\mu\nu} - h_{\mu\nu})\ .\autoeqno{nolog}
$$
Another description has been proposed by the present authors (see
Kijowski \& Magli 1992) in terms of the quantity
$$
S := \due \log K \autoeqno{log}
$$
where
$$
K^\mu_\nu := g^{\mu\lambda} (h_{\lambda \nu} - u_\lambda u_\nu )\ .
$$
Such strain tensors are both orthogonal to $u$ and vanish for the
locally
 relaxed state. There is a one-to-one correspondence between them.
Therefore, both descriptions are equivalent.
FFrom the theoretical point of view,
the description \eqref{log} is
somewhat preferable since $S$ is free to assume any value of a
symmetric
 tensor orthogonal to $u$, whereas $\Sigma$ is subject to a rather
involved matrix inequality ($h=E - 2\Sigma \geq 0$). In the
linearized
version of the theory, both descriptions obviously coincide.

The internal elastic energy
accumulated in each portion of the body
is a function of its deformation, described by its state of strain.
Hence, the energy depends upon both the
physical metric $g$ and the material metric $h$ via their
combination $S$
 (or $\Sigma$).

\pretitolo
{\bf 4. Carter-Quintana variational principle } \postitolo

A kinematically admissible $h$ may describe a real physical
situation
if and only if it satisfies also the dynamical equations of the
theory,
which can be derived from a variational principle. For this purpose
Carter and Quintana consider a bigger physical system, composed of
both
the elastic body and the gravitational field interacting with it.
The
total Lagrangian density of such a system is:
$$
\Lambda =-\menog \left(\frac 1{16\pi}R + \epsilon \right) \ , $$
where $\epsilon$ is the rest-frame energy density of the material
and
$R$ is the scalar curvature of the space-time. Keeping the material
configuration $h$ fixed and varying $\Lambda$ with respect to the
gravitational field $g$ we obtain Einstein equations $$ G_{\mu\nu}
=8\pi
T_{\mu\nu} \autoeqno{Einstein} $$ where $T_{\mu\nu}$ is the
symmetric
energy--momentum tensor: $$ T_{\mu\nu} := 2\diff{\epsilon}{\gia
\mu\nu}-\epsilon \gi \mu\nu \ . \autoeqno{deft}
$$
In this approach the dynamical equations for $h$ arise only as the
compatibility conditions
$$
\nadib \mu T^{\mu\nu} =0 \autoeqno{nt}
$$
of \eqref{Einstein} with the Bianchi identities, satisfied by the
Einstein tensor $G$.
Among eqs. \eqref{nt}, only three are independent (elasticity
theory
 has three degrees of freedom, as we have seen) and in fact in the
Carter
and Quintana paper it is shown that the equation $$ u_\nu \nadib
\mu
T^{\mu\nu} = 0
$$
holds identically. The proof is rather involved and requires the
introduction of the concept of {\it convected derivative}.

But the
fundamental drawback of the Carter-Quintana method of deriving
dynamical equations is that elastodynamics and geometro--dynamics
 can not
be separated. Indeed, the gravitational field can not be given {\it
a
priori}, as e.g.~in special relativity. The theory of ``test''
elastic
bodies in the flat Minkowski space is therefore excluded, because
the
space-time metric has always to be considered as a dynamical
variable.
Also the non-relativistic theory, which we know to be a Lagrangian
theory
(see e.g.~Sommerferld 1950) is automatically excluded. Moreover, it
is
very difficult to impose upon the general theory the existence of
symmetries in order to handle specific problems, as the equilibrium
of
axisymmetric solid stars (see e.g. Carter 1973, Carter \& Quintana
1975,
Quintana 1976, Priou 1992).

To illustrate the Carter-Quintana approach let us consider
classical,
Maxwell electrodynamics. Here, the field configurations are
described by
the electromagnetic, skew-sym\-met\-ric tensor $F_{\mu\nu}$. In our
example $F$ plays a role analogous to that of $h$ in
elastodynamics. The
first pair of Maxwell equations:
$$
\dib \mu F_{\rho\sigma} +
\dib \rho F_{\sigma\mu} +
\dib \sigma F_{\mu\rho} =0 \autoeqno{hm} $$
can be considered as the kinematical condition, analogous to
\eqref{lie}. We will show that
the remaining dynamical Maxwell equations can be obtained from the
general-relativistic
procedure, analogous to the Carter-Quintana approach. For this
purpose,
consider the
Lagrangian of the system composed of both the electromagnetic and
the
gravitational field:
$$
\Lambda =-\menog \left( \frac 1{16\pi}R + \frac 14 F_{\rho\sigma}
F^{\rho\sigma} \right) \ . \autoeqno{LA} $$
Keeping the electromagnetic field $F$ fixed and varying $\Lambda$
with
respect to the gravitational field $g$ we obtain Einstein
equations: $$
G^\mu_\nu =8\pi \left(
F^{\mu\rho}F_{\rho\nu}+\frac 14
F^{\sigma\rho}F_{\sigma\rho}\id \mu\nu \right)\ . $$
Now, Maxwell equations may be obtained as the compatibility
conditions
for the above system. Indeed, the Bianchi identities imply $$
F_{\rho\nu}\nadib \mu F^{\mu\rho}
+F^{\mu\rho}\nadib \mu F_{\rho\nu} +\due F^{\rho\sigma} \nadib \nu
F_{\rho\sigma}=0\ . $$
Using the skew--symmetry of $F$
and the kinematical
equations \eqref{hm}, it is easy to see that the last two terms
cancel.
Hence, for a generic, non singular $F$, we obtain the second pair
of
Maxwell equations $$
\nadib \mu F^{\mu\rho} =0\ .\autoeqno{nhm} $$
\par
Of course, the above approach to electrodynamics, although
equivalent to
 the standard one, is very inconvenient if one wants to describe
the
electromagnetic field in a given space-time geometry (e.g.~in
Minkowski
space).

\pretitolo
{\bf 5. Relativistic elasticity in terms of potentials.} \postitolo

To obtain the standard variational principle for the Maxwell field
(not
necessarily
coupled with the gravitational field), one introduces the
electromagnetic
 potential $A_\mu$, such that $$
F_{\mu\nu}=\dib \mu A_\nu -\dib \nu A_\mu\ . \autoeqno{pot} $$
In terms of the potential, electromagnetism becomes a gauge theory:

equations \eqref{hm} are automatically satisfied and equations
\eqref{nhm} become second--order dynamical equations for the
potentials.
Such equations can be obtained directly from the variation of the
electromagnetic Lagrangian with respect to the potentials. There is
no
need to couple electrodynamics to gravity, although it is possible.
In
the latter case the total Lagrangian is again equal to \eqref{LA}.

In the case of elastomechanics, we propose a similar approach.
As ``potentials'' of
the theory we take
the collection of all the idealized ``molecules'' of the material,
organized in an abstract 3--dimensional differential manifold $Z$,
called
the {\it material space}. The space--time configuration of the
material
is completely described if we specify a mapping ${\cal G }:{\cal M
} \to
Z \ $. To a given point of the physical space--time (i.e. to a
given
point of the space and a given instant of time) the mapping assigns
the
``molecule'' of the material which passes through that particular
point
at that particular time. Given a coordinate system $(\xi ^a)$
($a=1,2,3$)
in $Z$ and a coordinate system $(x^\mu )$ in ${\cal M}$, the
configuration of the material is described by three functions $\xi
^a
=\xi ^a(x^\mu )$, which play a role similar to that of the
potentials
$A_\nu =A_\nu (x^\mu)$ in electrodynamics.

The above potentials may be used for the description of any
continuous
material (see e.g.~Kijowski, Pawlik \& Tulczyjew 1979 and Kijowski,
Sm\'olski \& G\'ornicka 1990 for the description of
thermo-hydrodynamics). The particular case of a barotropic, elastic
material requires the use of a metric structure in the material
space
(Hernandez 1970, Glass \& Winicour 1972, Kijowski \& Magli 1992).
We
assume, therefore, that the space $Z$ is equipped with a riemannian
(positive) metric $\gamma _{ab}$, the {\it material metric}. This
metric
is frozen in the material and {\it is not} a dynamical object of
the
theory. It is given {\it a priori} for a given material, like e.g.
the
equation of state for a fluid. To understand its physical meaning
consider an infinitesimal portion of the material. This portion
will tend
spontaneously to a relaxed state when no external forces act on it.
The
metric $\gamma $ may now be defined as describing the rest-frame
space
distances between adjacent ``molecules'', measured in such a {\it
locally} relaxed state. There are, however, materials possessing no
{\it
globally} relaxed state. This happens if
$Z$ is not isometric to any hypersurface of ${\cal M}$, e.g. when
$Z$ is
 curved and ${\cal M}$ is flat. We see that materials
with internal stresses can also be described in this way. \par

Given the space--time configuration, we introduce the space-time
version
$h$ of the material metric as the geometric pull--back of the
metric
$\gamma$ from $Z$ to ${\cal M}$, i.e. $$ h := (d\xi )^* \gamma \ .
$$
In terms of coordinates we have:
$$
h_{\mu \nu }:=\gamma_{ab}\ixi a\mu \ixi b\nu \ ,\autoeqno{hmunu} $$
where we denote
$\ixi a\mu :=\partial _\mu \xi ^a $. The $3\times 4$ matrix
$\left(\partial _\mu \xi ^a \right) $ is called the {\it
relativistic
deformation gradient}.

Hence, similarly to electrodynamics, the physical field has been
expressed in
terms of the derivatives of the
potentials: formula \eqref{hmunu} in elastomechanics is analogous
to
formula \eqref{pot} in electrodynamics. It solves {\it
automatically}
the
kinematical condition \eqref{lie}, just as \eqref{pot} solves
automatically the kinematical condition \eqref{hm} in
electrodynamics.
Indeed, we have:
$$
\eqalign{
\L_{u} \h \mu\nu &=
u^\lambda \partial_\lambda \h \mu\nu +
\h \mu\lambda \partial_\nu u^\lambda + \h \nu\lambda \partial_\mu
u^\lambda =u^\lambda \left( \partial_\lambda \h \mu\nu -
\partial_\nu \h
\mu\lambda - \partial_\mu \h \nu\lambda \right) \cr &=
u^\lambda \left[ (\partial_\lambda \gamma_{ab}) \ixi a\mu \ixi b\nu
+
\gamma_{ab}\ixi a{\mu \lambda} \ixi b\nu +\gamma_{ab}\ixi a\mu \ixi
b{\nu
\lambda} \right. \cr & - (\partial_\nu \gamma_{ab}) \ixi a\mu \ixi
b\lambda - \gamma_{ab}\ixi a{\mu \nu} \ixi b\lambda -
\gamma_{ab}\ixi
a\mu \ixi b{\lambda \nu } \cr & \left. - (\partial_\mu \gamma_{ab})
\ixi
a\nu \ixi b\lambda - \gamma_{ab}\ixi a{\nu \mu} \ixi b\lambda -
\gamma_{ab}\ixi a\nu \ixi b{\lambda \mu } \right] \ . }
\autoeqno{ala} $$
To prove that the above expression vanishes identically, let us
first
observe that the deformation gradient is automatically orthogonal
to the
velocity vector:
$$
u^\mu \ixi a\mu \equiv 0 \ ,
$$
because $u$ is tangent to the trajectories of the material
molecules,
i.e to the lines given by the equation $\xi^a = \hbox{const}$. This
implies also:
$$
u^\lambda (\partial_\lambda \gamma_{ab}) = u^\lambda \ixi c\lambda
\frac {\partial \gamma_{ab}}{\partial \xi^c }
\equiv 0 \ ,
$$
and the other terms in \eqref{ala} cancel, which ends the proof.

The
theory is, of course, invariant with respect to reparametrizations
of
the material space, which play the role of gauge transformations.
Therefore, the fields
$\xi^a$ may be regarded as gauge potentials for the ``elasticity
field''
 $h$. They describe the three degrees of freedom of the system in
an
explicit way.
\par
The group of gauge transformations of the entire theory (elasticity

interacting with gravity)
is therefore the product of the group
of space-time diffeomorphisms (which is the gauge group of general
relativity) by the group of diffeomorphisms of the material space.

\pretitolo
{\bf 6. Dynamics}
\postitolo

The physical laws describing the mechanical properties of the
elastic
material can now
be formulated in terms of a system of second order hyperbolic
partial
differential equations for the 3 unknown fields $\xi ^a$. The
equations
can be obtained from the Lagrangian
$$\Lambda =-\menog \epsilon \ .\autoeqno{Lagr} $$
considered as a function of the fields $(\xi^a)$ and their first
derivatives. They assume the form of the Euler--Lagrange equations:
$$
\dib \mu \diff{\Lambda}{\ixi a\mu} -\diff{\Lambda}{\xi^a}
=0\ .\autoeqno{el}
$$

We are going
to prove the equivalence of the above equations with the equations
\eqref{nt} of Carter-Quin\-ta\-na.
For this purpose, we consider
the canonical
energy--momentum tensor of our theory:
$$
{\cal T}^\mu_\nu :=\frac 1\menog \left(
\diff \Lambda{\ixi a\mu}\ixi a\nu -\id \mu\nu \Lambda \right)\ .
\autoeqno{canonico}
$$
Due to the standard N\"other argument, the Euler-Lagrange field
equations imply the energy-momentum conservation:
$$
\nadib \mu {\cal T}^{\mu\nu} =0 \ . \autoeqno{conserv} $$
The same argument shows (see Kijowski \& Magli, 1992) that the
identity
$$
u_\nu \nadib \mu {\cal T}^{\mu\nu} \equiv 0 $$
holds automatically because of the gauge invariance of the theory.
Hence, in the case of elastomechanics there are only 3 independent
equations among the 4 equations \eqref{conserv}. To prove their
equivalence with the 3 Euler-Lagrange equations \eqref{el}, it is
now
sufficient to show that both energy-momentum tensors coincide up to
a
sign: $$ {\cal T}_{\mu\nu} \equiv - T_{\mu\nu} \ .\autoeqno{uguali}
$$ In
fact, this is a particular case of the general
Belinfante--Rosenfeld
theorem (Belinfante 1940, Rosenfeld 1940). The theorem uses the
invariance of the Lagrangian with respect to the space--time
diffeomorphisms. This means that it may be easily proved in
framework
which is more general than the one used so far in the present
paper,
namely, we are free to consider anisotropic elastic bodies as well.
In
order to introduce the variational principle for such bodies, we
define
the following tensor in the material space: $$
\te ab :=\gi \mu\nu \ixi a\mu \ixi b\nu\ , $$
{}from the geometrical point of view,
such a tensor is the ``image'' of the physical metric in the
material
space, i.e. $\te ab$ plays in $Z$ the same role played by
$h_{\mu\nu}$ in
${\it \cal M}$. For a general, possibly anisotropic body, we assume
the
energy density to be a function of the whole $\te ab$:
$$
\epsilon = \epsilon (\te ab ,\xi )\ .
$$
The particular case of isotropic bodies corresponds to an internal
energy which is a function of the invariants of $\te ab$ only;
of
course, the invariants of $\te ab $ coincide with the corresponding
 invariants of $h_{\mu\nu}$, like e.g.:
$$
\Theta^a_a =\gamma_{ab}\Theta^{ab} =\gamma_{ab}\ixi a\mu \ixi b\nu
\gia \mu\nu = h_{\mu\nu} \gia \mu\nu =h^\mu_\mu\ . $$
\par
According to the definition \eqref{canonico}, we have:
$$
-{\cal T}^\mu_\nu =
\diff \epsilon{\ixi a\mu}\ixi a\nu -\id \mu\nu \epsilon =
\diff \epsilon{\te cd}
\diff{\te cd}{\ixi a\mu} \ixi a\nu
-\epsilon \id \mu\nu =
2\diff \epsilon{\te ac}\gia \mu\alpha
\ixi a\alpha \ixi b\nu
-\epsilon \id \mu\nu
\ ,
$$
On the other hand, due to definition \eqref{deft}, we have $$
T_{\mu\nu} =
2\diff \epsilon{\gia \mu\nu} -\epsilon \gi \mu\nu = 2\diff
\epsilon{\te
ab}
\diff {\te ab}{\gia \mu\nu}
-\epsilon \gi \mu\nu=
2\diff \epsilon{\te ab}
\ixi a\mu \ixi b\nu
-\epsilon \gi \mu\nu
\ ,\autoeqno{cano}
$$
which ends the proof of the Belinfante--Rosenfeld identity
\eqref{uguali}.
It is worthwhile to note that the first term in \eqref{cano} is
automatically orthogonal to the velocity. Thus, the
energy--momentum
tensor assumes the canonical form $$ T_{\mu\nu}=\epsilon u_\mu
u_\nu
+p_{\mu\nu} $$ where the {\it pressure tensor} $p_{\mu\nu}$ is
orthogonal
to the velocity and is given by
$$
p_{\mu\nu} := 2
\diff \epsilon{\te ab}
\ixi a\mu \ixi b\nu
-\epsilon E_{\mu\nu}
\ .
$$

\pretitolo
{\bf 7. Concluding remarks}
\postitolo

In principle, the theory presented
here follows the lines of the approach to relativistic elasticity
proposed by Cattaneo (1973).
There is, however, a fundamental difference: Cattaneo describes the

configuration of the material always with respect to a fixed,
static
reference configuration. Choosing a space--like surface
$\{t={\rm const}\}$ of this reference configuration we obtain a
specific
 representation of our material space $Z$. We find it, however,
more
natural to use only the abstract, metric structure of $Z$, instead
of
working always with two different space-times: one to describe the
actual
configuration and another to describe the reference configuration.
Both
theories are equivalent in the particular case of flat materials
(no
internal--frozen stresses, flat material metric) and fixed
gravitational
field (e.g. special relativity). In order to study the interaction
between gravity and elasticity in Cattaneo's formulation, it is
necessary
to follow an {\it ad hoc} approach, which consists in defining a
reference state of the material when the gravity is hypothetically
``switched off'' (Newton constant set equal to zero), and then in
studying the evolution during a process of adiabatic restore of the
coupling constant to its own value (Cattaneo 1973, Cattaneo \&
Gerardi
1975). Mathematically, this is a nice way to overcome the
difficulties of
defining a reference state. From the physical point of view,
however,
switching off the gravity is poorly justified.\par An approach
similar to
ours has also been given by Maugin (1978). However, this author
works in
a ``direct'' picture rather than in the ``inverse'' picture,
proposed in
our paper. The difference between these two pictures consists in
inverting the role of the space parameters $(x^k)$ and the fields
$(\xi^a)$. In the ``direct'' picture the configuration of the
material is
described by the 3 fields $x^k$ depending upon 4 independent
parameters
$(x^0 , \xi^a )$. For this purpose a (3+1)-decomposition of the
space-time has to be chosen, which breaks the explicit relativistic
invariance of the theory. However, the theory proposed by Maugin
remains
relativistic-covariant and its results are independent of a
specific
choice of the (3+1)-decomposition.

The advantage of our approach consists in the fact that we really
eliminate all the constraints
in a fully relativistic-invariant way, reducing the number of
independent degrees of freedom to the three independent functions
$\xi^a$.
We hope this approach to be useful for specific applications like
e.g.
the description of the equilibrium and the oscillations of neutron
stars
crusts in a general relativistic context. Some steps in this
direction
have already been done (Magli \& Kijowski 1992, Magli 1993a,b).
Moreover,
as will be shown in a forthcoming paper (Kijowski \& Magli 1993),
our
formulation naturally leads to the hamiltonian version of the
theory, the
canonical variables being the fields $\xi^a$ and their conjugate
momenta
$\pi_a =\partial\Lambda /\partial{\dot \xi^a}$. The Poisson bracket
between them assumes its canonical, delta--like form. The above
hamiltonian structure is common for any (relativistic or
non-relativistic) mechanics of continuous media (see also Kijowski,
Sm\'olski \& G\'ornicka 1990 and Jezierski \& Kijowski 1991). It is
given
by a standard Legendre transformation, once the Lagrangian
\eqref{Lagr}
is expressed in terms of the (independent) variables $(\xi^a)$ and
their
first (unconstrained) derivatives (see also Kijowski \& Tulczyjew
1979).

\vfill\eject

\pretitolo
\centerline{\bf References}
\postitolo

\item{\ }
Belinfante, F. J. (1940)
On the current and the density of the electric charge, the energy,
the linear momentum and the angular momentum of arbitrary fields.
Physica {\bf 7} p. 449-474.\capo

\smallskip

\item{\ }
Carter, B., Quintana, H. (1972)
Foundations of general relativistic high pressure elasticity
theory.
Proc. R. Soc. Lond. A{\bf 331}, p. 57-83.\capo

\smallskip

\item{\ }
Carter, B. (1973a)
Speed of sound in a high-pressure general relativistic solid. Phys.
Rev.
{\bf D} 7, p. 1590-1593.\capo

\smallskip

\item{\ } Carter, B. (1973b)
Elastic perturbation theory in General Relativity and a variation
principle for a rotating solid star. Comm. Math. Phys. {\bf 30}, p.
261-280 (1973).\capo

\smallskip

\item{\ }
Carter, B., Quintana, H. (1975)
Stationary elastic
rotational deformation of a relativistic neutron star model. Ap. J.

{\bf 202}, p. 511-522.\capo

\smallskip

\item{\ }
Carter, B. (1980)
Rheometric structure theory, convective differentiation and
continuum
electrodynamics.
Proc. R. Soc. Lond. A{\bf 372}, p. 169-200.\capo

\smallskip

\item{\ }
Cattaneo, C. (1973)
Elasticit\'e Relativiste.
Symp. Math. {\bf 12}, Acad. Press NY, p. 337-352.\capo

\smallskip

\item{\ }
Cattaneo, C., Gerardi, A. (1974)
Su un problema di equilibrio elastico
in relativit\'a generale.
Rend. Mat. {\bf 8} p. 187-200.\capo

\smallskip

\item{\ }
Glass, E.N., Winicour, J. (1972)
General Relativistic elastic systems.
J. Math. Phys. {\bf 13} p.1934-1940.\capo

\smallskip

\item{\ }
Glass, E.N., Winicour, J. (1973)
A geometric generalization of Hooke's law. J. Math. Phys. {\bf 14}
p.1285-1290.\capo

\smallskip

\item{\ }
Hernandez, W.C. (1970)
Elasticity in general relativity.
Phys. Rev. D {\bf 1}, p. 1013-1018.

\smallskip

\item{\ }
Jezierski, J., Kijowski, J. (1991)
Thermo-hydrodynamics as a field theory.
In {\it Hamiltonian Thermodynamics}, editors S. Sieniutycz and P.
Salamon, Taylor and Francis Publishing Company.\capo

\smallskip

\item{\ }
Kijowski, J., Magli, G. (1992)
Relativistic elastomechanics as a lagrangian field theory. J. Geom.

Phys. {\bf 9}, p. 207-223.\capo

\smallskip

\item{\ }
Kijowski, J., Magli, G. (1993)
Hamiltonian formulation of general relativistic elasticity theory.
 In preparation.\capo

\smallskip

\item{\ }
Kijowski, J., Pawlik, B., Tulczyjew W.M. (1979) A variational
formulation of non-gra\-vi\-ta\-ting and gravitating hydrodynamics,

Bull.
Acad. Polon. Sci. {\bf 27} p. 163-170.\capo

\smallskip

\item{\ }
Kijowski, J., Sm\'olski, A., G\'ornicka A. (1990) Hamiltonian
theory
of self--gravitating perfect fluids and a method of effective
deparametrization of Einstein theory of gravitation. Phys. Rev. D
{\bf
41}, p. 1875--1884.\capo

\smallskip

\item{\ }
Kijowski, J., Tulczyjew W.M. (1979)
{\it A symplectic framework for field theories}. Lecture Notes in
 Physics {\bf 107}, Springer, Berlin.

\smallskip

\item{\ }
Magli, G., Kijowski, J. (1992)
A generalization of the relativistic equilibrium equations for a
non
 rotating star. Gen. Rel. Grav. {\bf 24}, p. 139-158.\capo

\smallskip

\item{\ }
Magli, G. (1993a)
The dynamical structure of the Einstein equations for a non
rotating
star. Gen. Rel. Grav. {\bf 25}, p. 441-460.\capo

\smallskip

\item{\ }
Magli, G. (1993b)
Axially symmetric, uniformly rotating neutron stars in General
Relativity: a non--perturbative approach Gen Rel. Grav. {\bf 25},
p.
1277-1293\capo

\smallskip

\item{\ }
Maugin, G.A. (1971)
Magnetized deformable media in General Relativity. Ann. Inst. Henri

Poincar\'e A {\bf 15}, 2, p.275-302.\capo

\smallskip

\item{\ }
Maugin, G.A. (1977)
Infinitesimal discontinuities in Initially Stressed Relativistic
Elastic
Solids.
Comm. Math. Phys. {\bf 53}, p. 233-256.\capo

\smallskip

\item{\ }
Maugin, G.A. (1978)
On the covariant equations of the relativistic electrodynamics of
continua III. Elastic solids. J. Math. Phys. {\bf 19}, p.
1212-1219.
\capo
\smallskip

\item{\ }
Maugin, G.A. (1978)
Exact Relativistic theory
of wave propagation in prestressed nonlinear elastic solids. Ann.
Inst.
 Henri Poincar\'e A {\bf 28}, p.155-185.\capo

\smallskip

\item{\ }
Mc Dermott, P.N., Van Horn, H.N., Hansen, C.J. (1988) Nonradial
oscillations of neutron stars. Ap. J. {\bf 325}, p. 725--748.\capo

\smallskip

\item{\ }
Priou, D. (1992)
The perturbations of a fully general
relativistic and rapidly rotating neutron star - I. Equations of
motion
 for the solid crust. Mon. Not. R. Astr. Soc. {\bf 254}, p. 435
(1992).\capo

\smallskip

\item{\ }
Quintana, H. (1976)
The structure equations
of a slowly rotating, fully relativistic solid star. Ap. J. {\bf
207},
 p. 279-288.\capo

\smallskip

\item{\ }
Rosenfeld, L. (1940)
Sur le tenseur d'impulsion-energie.
Acad. Roy. Belg. {\bf 18}, p. 1--30.\capo

\smallskip

\item{\ }
Sommerfeld, A. (1950) {\it Mechanics of deformable bodies}, Acad.
Press N.Y. \capo

\bye